\documentclass[aps,pra,superscriptaddress, twocolumn]{revtex4}
\usepackage[intlimits]{amsmath}
\usepackage{amsfonts}
\usepackage{psfrag}
\usepackage{subfigure}
\usepackage[usenames]{color}
\usepackage{multirow}
\usepackage{braket}
\usepackage{epstopdf}
\usepackage{tikz}

\usepackage{ifpdf}

\ifpdf
\usepackage{epstopdf}
\usepackage[pdftex]{hyperref}
\else
\usepackage[hypertex,ps2pdf]{hyperref}
\fi

\advance\voffset by -1cm
\advance\hoffset by -0.5cm
\newcommand\be            {\begin{equation}}
\newcommand\bea           {\begin{equation}\begin{array}l\displaystyle}
\newcommand\ee            {\end{equation}}
\newcommand\bes           {\begin{subequations}}
\newcommand\esu           {\end{subequations}}

\newcommand{\bigx}[1]{\bBigg@{#1}}

\newcommand\eps           {\varepsilon}

\newcommand\p            {\partial}

\def\3pt#1#2#3{{\langle{#1}\vert{#2}\vert{#3}\rangle}}

\newcommand\doi[2]        {\href{http://dx.doi.org/#1}{#2}}

\begin{document}

\title{Universal off--diagonal long--range order behaviour \\
for a trapped Tonks--Girardeau gas}

\author{A. Colcelli}
\affiliation{SISSA and INFN, Sezione di Trieste, Via Bonomea 265, I-34136 
Trieste, Italy}
\author{J. Viti}
\affiliation{International Institute of Physics \& ECT, UFRN, 
Campos Universit\'ario, Lagoa Nova 59078-970 Natal, Brazil}
\author{G. Mussardo}
\affiliation{SISSA and INFN, Sezione di Trieste, Via Bonomea 265, I-34136 
Trieste, Italy}
\author{A. Trombettoni}
\affiliation{CNR-IOM DEMOCRITOS Simulation Center, Via Bonomea 265, I-34136 
Trieste, Italy} 
\affiliation{SISSA and INFN, Sezione di Trieste, Via Bonomea 265, I-34136
Trieste, Italy}

\begin{abstract}
\noindent
The scaling of the largest eigenvalue $\lambda_0$ of the one--body density matrix of a system with respect to its particle number $N$
defines an exponent $\mathcal{C}$ and a coefficient $\mathcal{B}$ via the asymptotic relation $\lambda_0 \sim \mathcal{B}\,N^{\mathcal{C}}$.
The case $\mathcal{C}=1$ corresponds to off--diagonal long--range order. For a one--dimensional homogeneous Tonks--Girardeau gas, 
a well known result also confirmed by bosonization gives instead $\mathcal{C}=1/2$. Here we investigate the inhomogeneous case, initially addressing 
the behaviour of $\mathcal{C}$ in presence of a general external trapping potential $V$. We argue that the value $\mathcal{C}= 1/2$ characterises the hard--core system 
independently of the nature of the potential $V$. We then define the exponents $\gamma$ and $\beta$ which describe the scaling with $N$ of the peak of 
the momentum distribution and the natural orbital corresponding to $\lambda_0$ respectively, and we derive the scaling relation $\gamma + 2\beta= \mathcal{C}$. 
Taking as a specific case the power--law potential $V(x)\propto x^{2n}$, we give analytical formulas for $\gamma$ and $\beta$ as functions of $n$. Analytical predictions 
for the coefficient $\mathcal{B}$ are also obtained. These formulas are derived exploiting a recent field theoretical formulation and checked against numerical results. The agreement is excellent. 

\end{abstract}
\maketitle

\maketitle
\section{Introduction}
\label{Introduction}
\noindent
The one--body density matrix (OBDM) $\rho(x,y)$ is a quantity of central importance for the statistical properties of interacting quantum systems.
In a second quantized formalism for a many-body bosonic system it can be written as the one--particle correlation function 
\be
\rho(x,y)\,=\,\left  \langle\Omega|  \Psi^\dagger(x) \Psi(y) |\Omega\right \rangle\,,
\ee
where $\Psi$ and $\Psi^\dagger$ are bosonic field operators and $|\Omega\rangle$ is the many-body ground state.
The eigenvalues $\lambda_j$ of the OBDM
are defined by the solution of the integral equation 
\be
\label{eigen_eq}
\int dy~\rho(x,y)\,\varphi_j(y)=\,\lambda_j\,\varphi_j(x)\,\,\,, 
\ee
where the functions $\varphi_j$ are usually called natural orbitals \cite{Pitaevskii16}. The scaling of the largest eigenvalue $\lambda_0$ of (\ref{eigen_eq}) with respect to the total number $N$ of particles gives information on whether the system exhibits Off--Diagonal Long--Range Order (ODLRO) and, as a consequence of the Penrose--Onsager criterion, Bose--Einstein condensation \cite{Penrose56,CNYang1962}. Indeed, in presence of ODLRO, the OBDM has non-vanishing off--diagonal elements, which implies for a homogeneous system a Dirac delta peak in the momentum distribution. This means that one has a macroscopic occupation of the lowest energy state, making therefore $\lambda_0$ scale with $N$, \textit{i.e.} $\lambda_0=O(N)$. On the other hand, when all the eigenvalues of $\rho(x,y)$ in Eq.~\eqref{eigen_eq} are order one, the system exhibits fermionic behaviour,  
obeying Pauli exclusion principle. 

Intermediate situations may arise however in 1D systems: Even though in such systems quantum fluctuations strongly deplete (and in the thermodynamic limit completely prevent) the Bose-Einstein condensate, the lowest eigenvalue of the OBDM scales nevertheless in a non-trivial way with respect to $N$, being not order one.
When $N$ is large, it is possible to define an exponent $\mathcal{C}$ via the relation
\be
\label{primaryDef}
\lambda_0 \sim \,\mathcal{B} N^\mathcal{C}\,.
\ee
Continuous variations of the exponent $\mathcal{C}$ give rise to a whole spectrum of possible order types -- long-range order, mid-range order and short-range order -- as discussed in  \cite{Yukalov91}. For one-dimensional bosons with a two--body delta--interaction, \textit{i.e.} the Lieb--Liniger model \cite{LiebLiniger}, and in absence of an external one--body trapping potential $V(x)$, the dependence of the exponent $\mathcal{C}$ on the interactions strength and the density of particles has been studied in \cite{CMT2018}. In this homogeneous case, denoting by $\gamma$ the coupling constant of the Lieb--Liniger model, in the weakly--interacting limit $\gamma \to 0$ we have $\mathcal{C} \to 1$, while in the strong--coupling regime $\gamma \to \infty$ we have $\mathcal{C} \to 1/2$, both results in agreement with bosonization \cite{Haldane,GiamarchiBook}. The limit of infinite $\gamma$ corresponds to the Tonks--Girardeau (TG) gas \cite{Girardeau1960}, \textit{i.e.} 1D hard--core bosons, and the result $\mathcal{C} = 1/2$ \cite{Lenard1964,Forrester2003} confirms the nature of the TG gas as intermediate between bosons and fermions, despite the fact that one-point observables are identical to those of fermions \cite{DasGirardeau2002}.

The definition of ODLRO, related to the long--distance behaviours of the OBDM, is of course valid both for homogeneous ($V(x)=0$) and inhomogeneous ($V(x) \neq 0$) systems. However,  beside the inherent difficulty of dealing with interacting systems (which is also present for the homogeneous systems), the study of the large-$N$ limit in presence of an external trapping potential shows several additional difficulties due to the lack of translational invariance. For instance: \textit{i)} numerical methods working at small $N$ may not be able to give the correct large-$N$ behaviour; \textit{ii)} the presence of the potential $V$ may spoil the validity of methods which explicitly exploit the translational invariance of the system, such as the perturbative expansions done in terms of Feynman diagrams in momentum space; \textit{iii)} for 1D systems, the external trapping potential typically also breaks the integrability of the homogeneous limit. Although one could derive useful information from approaches based on local density approximation, a complete study of the ODLRO behaviour 
of strongly correlated quantum systems in presence of external trapping potentials remains a challenging task.

With this main motivation, in this paper we focus on the characterization of the ODLRO in a TG gas at $T=0$ in the presence of external trapping potentials. The interest in such a 
study relies both on experimental and theoretical sides. Indeed, several progresses have been done in realising the TG gas with ultracold atoms and characterising its properties, such as momentum distribution and ground state energy \cite{Paredes2004,Kinoshita2004}. For the TG gas, one can also study the confinement of induced resonances as well as the crossover to the so-called super-TG gas \cite{Haller09}; in presence of a periodic potential, one can also address the quantum phase transition which induces to a Mott insulating state \cite{Haller10,Boeris16}. From a theoretical perspective, one is able to work out analytical results for this system thanks to its integrability \cite{Korepin,Franchini} and the Bose--Fermi equivalence~\cite{GirardeauHFmapping, GirardeauWright2000}, which permits to map the TG gas into a system of non--interacting spinless fermions. It is also known that for a trapped TG gas one can write a closed expression for the OBDM~\cite{PezerBuljan}. For all these reasons there is a broad interest in the study of correlation functions of the TG gas in different potentials, such as harmonic traps \cite{Lapeyre,Forrester2003,
MinguzziGangardt2005,Vignolo13,Collura13,Lang17,Rizzi18}, optical lattices \cite{Rigol2004,WeiGuLin,Lelas12,WangWangLi,Cartarius15,Astrakharchik16,Astrakharchik17}, 
disordered potentials \cite{Radic10,Seiringer16,Settino17}, or also of the super-TG state 
\cite{Astrakharchik05,Batchelor05,Kormos11,Girardeau12,Panfil13}
(see~\cite{ForresterCommunications2003} for additional references). 

In presence of a trapping potential, there are two ways in which one can take the large-$N$ limit: {\it a)} increasing $N$ and at the same time varying the parameters of the external potential $V$ ({\it e.g.}, the harmonic oscillator length for the harmonic potential) in such a way to keep fixed the density at the center of the trap, for instance; 
{\it b)} or fixing the parameters of the external potential $V$ and simply increasing $N$. It turns out that the scaling of the largest eigenvalue of the OBDM is the same in both cases (see Sec.~\ref{sub:pred_B} for more details).

In this paper we first show that the result ${\mathcal C}=1/2$ holds for a TG gas independently of the external potential $V$. We will argue that the universality of this result can be predicted by exploiting the expression of the OBDM of a TG gas in a generic trapping potential obtained in~\cite{BrunDubail}. Such an expression that is valid for large $N$, was derived expanding a Conformal Field Theory (CFT) approach introduced in~\cite{Calabrese2017}. 

We then discuss the emergence of other power--law behaviours for the Tonks-Girardeau gas. Suitably rescaling the space coordinate $x$, as explained in detail in Sec.~\ref{sec5},
one can study how the peak at zero momentum of the dimensionful momentum distribution $\tilde{\rho}(k)$ and the natural orbital $\varphi_0$ corresponding to $\lambda_0$, scale 
with the particle number. These two quantities define respectively two exponents, denoted by $\gamma$ and $\beta$, which we will show are related via $\gamma + 2 \beta = \mathcal{C}$. For simplicity, we will mainly refer to power--law potentials of the form $V(x) \propto x^{2n}$, interpolating between the harmonic potential ($n=1$) and the hard wall trap ($n\to\infty$). For these power--law potentials we are able to predict the dependence on $n$ of both $\gamma$ and $\beta$. We are also able to obtain accurate predictions for the coefficient $\mathcal{B}$ that appears as a pre-factor in the scaling relation~\eqref{primaryDef}. We corroborate these results using both a WKB approximation and direct numerical calculations. Finally we obtain the scaling with the particle number of the dimensionful momentum distribution peak. The latter is characterised by the same exponent $1/2$, irrespectively of the external potential, in analogy with the result for the largest eigenvalue $\lambda_0$. 

The paper is organised as follows. In Sec.~\ref{sec2} we revisit the OBDM of a TG gas and derive its expression in terms of the single--particle wave-functions of the system~\cite{PezerBuljan}. In Sec.~\ref{sec3}, we discuss the scaling behaviour for the largest eigenvalue of the OBDM and the peak of the momentum distribution, as well as a relation among these exponents. In Sec.~\ref{sec5} we present a numerical study of the scaling of $\lambda_0$ and $n_{\text{peak}}$ 
with respect to $N$: Our main results are reported in Tab.~\ref{table:finalresults}.  We finally gather our conclusions in Sec.~\ref{sec:Concl}, while some details about the numerical methods are reported in Appendix \ref{app:A}.

\section{The model and its one--body density matrix}
\label{sec2}
\noindent 
In the limit of infinite coupling, the Lieb--Liniger model  reduces to a system of $N$ impenetrable bosons  of mass $m$. Such a system is known as the TG gas \cite{Girardeau1960},  and it is described by the Schr\"odinger equation
\be 
H \,\psi(x_1,\dots,x_N)\,=\,E \,\psi(x_1,\dots,x_N)\,,
\label{eigenTGH}
\ee
where the Hamiltonian is written as a sum of single--particle Hamiltonians
\be
\label{HAM_TG}
H=\sum_{i=1}^N\left[-\frac{\hbar^2}{2m} \frac{\p^2}{\p x_i^2} + V(x_i)\right]\,.
\ee
In (\ref{eigenTGH}) the many-body wave-functions $\psi$ are symmetric in the exchange of two coordinates due to the bosonic statistics, although they vanish when two arguments have the same value for the hard--core interactions
\be
\label{eigen_null}
\psi|_{x_i = x_j}\,=\,0\,, \,\, \, \, \, \forall i\not =j=1,\dots,N.
\ee
The many--body wave-functions of the system can then be written in a Slater determinant form by adding sign functions to ensure the 
correct symmetry under coordinate exchange
\begin{multline}
\label{TG_mapping}
\psi(x_1,\dots,x_N)\,=\,\frac{1}{\sqrt{N!}} \det\left[\phi_k (x_l)\right]_{\substack{k=0,\dots,N-1,\\l=1,\dots,N}}\\
\prod_{1\le i<j\le N} \text{sgn}(x_i -x_j)\,. 
\end{multline}
This is the content of the well known Fermi--Bose equivalence \cite{GirardeauHFmapping, GirardeauWright2000}, where $\phi_k(x)$ denotes the $k$-th 
eigenfunction of the single--particle Schr\"odinger equation ($k=0,\dots,N-1$)
\be
\label{SP_schro}
\left[-\frac{\hbar^2}{2m} \frac{d^2}{d x^2} + V(x)\right]\phi_k(x)\,=\,\eps_k \phi_k(x)\,,
\ee
and $\eps_k$ is the corresponding single--particle energy.

The Hermitian OBDM $\rho(x,y)$ of the 1D quantum gas is defined as
\begin{multline}
\label{OBDM_def}
\rho(x,y)\,=\,N\int \prod_{i=2}^N~dx_i~\psi^{*}(x,x_2,\dots,x_N)\\\times \psi(y,x_2,\dots,x_N)\,.
\end{multline}
Notice that $\rho(x,y)$ is also often referred to in literature (for instance~\cite{Giam_rev}) as $g_1(x,y)$.
In the following, the integrals are meant to be between $-\infty$ and $+\infty$ each time that their extremes are not explicitly written. 

The solutions of the eigenvalue equation for the OBDM, \textit{i.e.} Eq.~\eqref{eigen_eq}, involve the natural orbitals $\varphi_j(x)$: They represent the effective single--particle states of the system, while the $\phi_k(x)$ can be viewed as the natural orbitals for the ideal fermionic gas \cite{GirardeauTriscari}. The natural orbitals are chosen to be orthonormal,  
$\int dx \,\varphi_i^*(x) \varphi_j(x)=\delta_{ij}$. The occupation numbers of the levels $j$, expressed by $\lambda_j$, satisfy the normalization condition
\be
\label{normcond}
\sum_{j} \lambda_j \,=\,N\,,
\ee
that is a consequence of $\int dx~\rho(x,x)=N$. Substituting Eq.~\eqref{TG_mapping} into Eq.~\eqref{OBDM_def} and expanding the Slater determinants along the first column, we obtain  
\begin{multline}
\label{OBDM_1}
\rho(x,y)\,=\,\frac{1}{(N-1)!} \sum_{i,j=0}^{N-1} (-1)^{i+j} \, \phi_i(y) \,\phi_j^{*}(x)\\ 
\times \int dx_2\hdots \int dx_{N}\det[f_k(x_r)]_{\substack{k=0,\dots,N-1, k\neq i\\r=2,\dots,N}}\\
\times \det[g_l(x_r)]_{\substack{l=0,\dots,N-1, l \neq j\\r=2,\dots,N}}\,,
\end{multline}
where $f_k(x_r)=\phi_k(x_r)\,\text{sign}(x-x_r)$, $g_l(x_r)=\phi_l^*(x_r)\,\text{sign}(y-x_r)$, with the index $k\not=i$ in the first determinant while $l\not=j$ in the second. It is worth to recall \textit{Andr\'eief formula}~(see~\cite{Forrester2018} for a nice recent historical note) 
\begin{multline}
\int dx_1\dots\int dx_N \det[f_{j}(x_k)]_{j,k=1}^N\det[g_{j}(x_k)]_{j,k=1}^N\\=N!\det\left[\int dx~ f_j(x) g_{k}(x)\right]_{j,k=1,\dots N}
\end{multline}
to transform the product of two determinants in Eq.~\eqref{OBDM_1} into the determinant of the product. It follows
\begin{multline}
\label{OBDM_2}
\rho(x,y)\,=\sum_{i,j=0}^{N-1} (-1)^{i+j} \, \phi_i(y) \,\phi_j^{*}(x)\\
\times\det\left[\int dt~f_k(t)g_l(t)\right]_{k\neq i, l\neq j}\,.
\end{multline}
We then substitute back in Eq.~\eqref{OBDM_2} the form for the functions $f_k$ and $g_l$ in terms of the single--particle wave-functions $\phi_k$ which are solutions of Eq.~\eqref{SP_schro}. Assuming $x>y$ and using the orthonormality condition $\int dt \, \phi_l(t) \, \phi_k^{*}(t)\,=\, \delta_{k,l}\,$, we obtain a compact form for the OBDM of a TG gas in a generic external potential $V(x)$  as
\begin{multline}
\label{OBDM_final}
\rho(x,y)\,=\,\sum_{i,j=0}^{N-1} (-1)^{i+j}\,\phi_i(y) \, \phi_j^{*}(x)\\
\times\det\left[\delta_{k,l} -2 \int_{y}^{x} dt\, \phi_l(t) \, \phi_k^{*}(t) \right]_{k\neq i, l\neq j}\,.
\end{multline}
Consider now  the matrix $P$ with  entries $P_{ij} = \delta_{i,j} - 2 \int_{y}^{x} dt\,\phi_j(t)\, \phi_i^{*}(t)$, for $i,j = 0,\dots,N-1$.  From Cramer theorem, one can check that Eq.~\eqref{OBDM_final} is actually equivalent to~\cite{PezerBuljan}
\be
\label{PezerBuljanFormula}
\rho(x,y)\,=\,\det(P)\sum_{i,j=0}^{N-1} \phi_i^{*} (x)\,[P^{-1}]_{ji} \,\phi_j (y)\,,
\ee
having again assumed $x>y$ without loss of generality.

In the following, for numerical computations involving the OBDM, we find simpler to use its expression given in  
Eq.~\eqref{OBDM_final}, which does not require explicitly the inverse of the matrix $P$. 
This expression also provides a non-trivial check of the large-$N$ limit derived in~\cite{BrunDubail} as we are now going to illustrate.

\section{Scaling of $\lambda_0$ and the momentum distribution peak}
\label{sec3}
\noindent 
In this Section we derive our predictions for the scaling of the largest eigenvalue of the OBDM and the momentum distribution peak of a TG gas in a generic external potential by using CFT within a semiclassical framework. In this Section we consider a generic trapping potential, assuming that its single--particle wave-functions $\phi_j(x)$ and its natural orbitals $\varphi_j(x)$ decay fast enough at large distances. In Sec.~\ref{sec5} we will focus on the case of an external power--law potential.

\subsection{One--Body Density Matrix in the semiclassical (CFT) limit}
\noindent
In the recent article \cite{BrunDubail}, Brun and Dubail studied the large distance behaviour of the OBDM of a TG gas in a generic trapping potential and in the semiclassical limit $\hbar \rightarrow 0$, by using CFT arguments coming from a previous analysis \cite{Calabrese2017}. The results of \cite{BrunDubail} were then extended in \cite{Brun18} to study the large distance behaviour of correlation functions of a Lieb--Liniger gas in a trap for arbitrary values of the coupling strength. Let's first briefly remind the framework and the main results of ref.~\cite{BrunDubail}. The semiclassical limit for the TG gas considered in~\cite{BrunDubail} is defined as
\begin{equation}
\label{slimit}
\hbar\rightarrow 0,\qquad \text{with}~m,~ V(x),~\mu~\text{fixed},
\end{equation}
where $\mu$ is the chemical potential. In the limit \eqref{slimit}, the inhomogeneous particle density can be obtained exactly within a local density approximation as
\begin{equation}
\rho(x)\,=\,\frac{1}{\pi\hbar}\sqrt{2m\,\left[\mu-V(x)\right]}. 
\end{equation}
In the following we denote by $x_1$ and $x_2$ (with $x_{2}>x_1$) the two solutions of the equation $\mu-V(x)=0$ and we assume that these are the only two solutions of this equation. For $x>x_2$ or $x<x_1$ the gas density is zero and the latter is effectively confined in a spatial region $x\in[x_1,x_2]$. The total number of particles in the system is
\be
\label{NDubail}
N\,=\,\int_{x_1}^{x_2} \rho(x) \,dx\,=\,\frac{1}{\pi \hbar} \int_{x_1}^{x_2} \sqrt{2m\,\left[\mu-V(x)\right]}\,dx\,.
\ee
It follows that the $\hbar\rightarrow 0$ limit is actually the thermodynamic limit $N\rightarrow\infty$ and this gives rise to the Thomas-Fermi approximation 
\cite{Pethick08}. To keep track of the leading $N$-dependence in the limit \eqref{slimit}, it is sufficient to observe that Eq.~\eqref{NDubail} implies 
\begin{equation}
\label{Ndeep}
N\hbar=\text{const},
\end{equation}
\textit{i.e.} $N=O(\hbar^{-1})$.  The main result of \cite{BrunDubail} is an expression for the OBDM of the TG gas in Eq.~\eqref{OBDM_final} in the limit \eqref{slimit} that is valid as long as $|x-y|\rho_{\max}\gg 1$, where $\rho_{\max}$ is the maximum density in the trap. Such an expression is
\be
\label{Dubailformula}
\rho_{\text{cft}}(\tilde{x},\tilde{y})\,=\sqrt{\frac{m}{2\hbar \widetilde{L}}}\frac{|C|^2\left | \sin\left(\frac{\pi \widetilde{x}}{\widetilde{L}} \right)\right |^{\frac{1}{4}}\,\left | \sin\left(\frac{\pi \widetilde{y}}{\widetilde{L}} \right)\right |^{\frac{1}{4}}}{\left | \sin\left(\frac{\pi }{\widetilde{L}} \frac{\widetilde{x}-\widetilde{y}}{2} \right)\right |^{\frac{1}{2}}\, \left | \sin\left(\frac{\pi }{\widetilde{L}} \frac{\widetilde{x}+\widetilde{y}}{2} \right)\right |^{\frac{1}{2}}}\,,
\ee
where $\left | C\right |^2$ is a numerical coefficient which can be expressed in terms of Barnes function $G(z)$ as $\left | C \right |^2 = \frac{G^{4}(3/2)}{\sqrt{2 \pi}}$, $\widetilde{L}$ is the time needed by a signal travelling with velocity $v$ to cover the interval $[x_1, x_2]$. The signal velocity $v(x)$ depends on the position $x$ as 
\be
\label{velocityDubail}
v(x)\,=\,\sqrt{\frac{2}{m}\,\left[\mu-V(x)\right]}\,.
\ee
One has then
\be
\label{LDubail}
\widetilde{L}\,=\,\int_{x_1}^{x_2} \frac{du}{v(u)}\,. 
\ee
In Eq.~\eqref{Dubailformula} $\widetilde{x}(x)$ represents the time needed to a signal emitted in $x_1$, with velocity
\eqref{velocityDubail}, to reach $x$, \textit{i.e.}
\be
\label{xtildeDubail}
\widetilde{x}(x)\,=\,\int_{x_1}^{x} \frac{du}{v(u)}\,. 
\ee
It should be noticed that in the limit \eqref{slimit} the condition $|x-y|\rho_{\max}\gg 1$ is satisfied up to distances $|x-y|=O({N}^{-1})$. To analyse Eq.~\eqref{eigen_eq} in the $\hbar\rightarrow 0$ limit, we can then safely replace $\rho(x,y)$ with Eq.~\eqref{Dubailformula} and restrict the integration domain to $y\in[x_1,x_2]$. Changing integration variable to $\tilde{y}(y)$ through Eq.~\eqref{xtildeDubail}, we obtain the semiclassical limit of Eq.~\eqref{eigen_eq} for the largest eigenvalue of the OBDM
\begin{equation}
\label{semi_eigen}
\int_{0}^{\tilde{L}}d\tilde{y}~w(\tilde{y})\rho_{\text{cft}}(\tilde{x},\tilde{y})\varphi_0(\tilde{y})=\lambda_0\varphi_0(\tilde{x}),
\end{equation} 
with $w(\tilde{y})=1/v(y(\tilde{y}))$. Plugging Eq.~\eqref{Dubailformula} into Eq.~\eqref{semi_eigen}, we observe that the limit $\hbar\rightarrow 0$ is consistent on both sides only if $\lambda_0=O(\hbar^{-1/2})$. Recalling Eq.~\eqref{Ndeep}, immediately we conclude that for $N\rightarrow\infty$
\begin{equation}
\label{universalresult}
\lambda_0\sim\mathcal{B} N^{1/2},
\end{equation}
namely, in the limit \eqref{slimit}, the scaling exponent $\mathcal{C}$ in Eq.~\eqref{primaryDef} is $1/2$, independently on the shape of the potential. The result $\mathcal{C}=1/2$, was found in the specific case of the harmonic potential in \cite{Forrester2003}. The numerical pre-factor $\mathcal{B}$ is instead potential-dependent and can be also explicitly calculated; we provide an example of such a computation in Sec.~\ref{sub:pred_B}. In Sec.~\ref{sub:num_results} we also support numerically the validity of Eq.~\eqref{universalresult} for different potentials, both increasing $N$ and keeping fixed the density. We will also estimate the value of the pre-factor $\mathcal{B}$ directly from Eq.~\eqref{OBDM_final}, thus providing another non-trivial check of Eq.~\eqref{Dubailformula}, which, it is worth stressing again, was derived relying on field theoretical arguments only.

A way to understand the validity of the result (\ref{universalresult}) for a generic potential $V$, consists of observing that the TG gas is the strong interacting limit of the Lieb--Liniger model. Writing the Lieb--Liniger Hamiltonian in the homogeneous case as $H_{LL}=-\frac{\hbar^2}{2m} \sum_{i} \frac{\p^2}{\p x_i^2} + 2 \lambda \sum_{i<j} \delta(x_i-x_j)$, the coupling constant $\gamma$ is defined by $\gamma=2m\lambda/\hbar^2 \rho$, where $\rho$ is the density and the TG gas is obtained when $\lambda \to \infty$. When, on the contrary, the system is inhomogeneous for the presence of the external potential $V$, then $\rho$ becomes space-dependent $\rho \to \rho(x)$, but notice that for $\lambda \to \infty$ one has again 
$\gamma \to \infty$, from which one can argue that the result (\ref{universalresult}) should continue to hold.

Eq.~\eqref{universalresult} is intended to describe the scaling of $\lambda_0$ when the shape of the external potential is fixed and one varies $N$. We will see from numerical calculations that the same power--law scaling for $\lambda_0$ emerges when the density of particles in the external potential is fixed and one varies $N$ and the trap parameters accordingly.

\subsection{Momentum Distribution}
\noindent 
We consider here the small--$k$ behaviour of the momentum distribution $\tilde{\rho}(k)$ of the system, defined as
\be
\label{mom_distr_def}
\tilde{\rho}(k)\,=\,\frac{1}{2\pi} \int
dx \int
dy \, \rho(x,y) e^{-i\,k\,(x-y)}\,.
\ee
From (\ref{eigen_eq}) we have the eigendecomposition
\be
\rho(x,y) \,=\,\sum_j \lambda_j\, \varphi_j^*(y) \, \varphi_j(x)\,,
\ee
that, substituted into Eq.~\eqref{mom_distr_def}, gives 
\be
\label{mom_distr_occnumb}
\tilde{\rho}(k)\,=\,\sum_j \lambda_j \lvert \tilde{\varphi}_j(k) \rvert^2\,,
\ee
where $\tilde{\varphi}_j(k)=\frac{1}{\sqrt{2\pi}}\int dx~e^{-ikx}\varphi_j(x)$ is the Fourier transform of the natural orbital. 
Hence the zero-momentum distribution $n_{\text{peak}}\equiv\tilde{\rho}(k=0)$ is given by 
\be
\label{n(k=0)}
n_{\text{peak}}\,=\,\frac{1}{2 \pi} \sum_j \lambda_j \,M_j\,.
\ee
where the quantities $M_j \, \equiv \,\left | \int dx~\varphi_j(x)\right|^2$ involve the natural orbitals. 

\begin{figure}[t]
\includegraphics[width=\columnwidth]{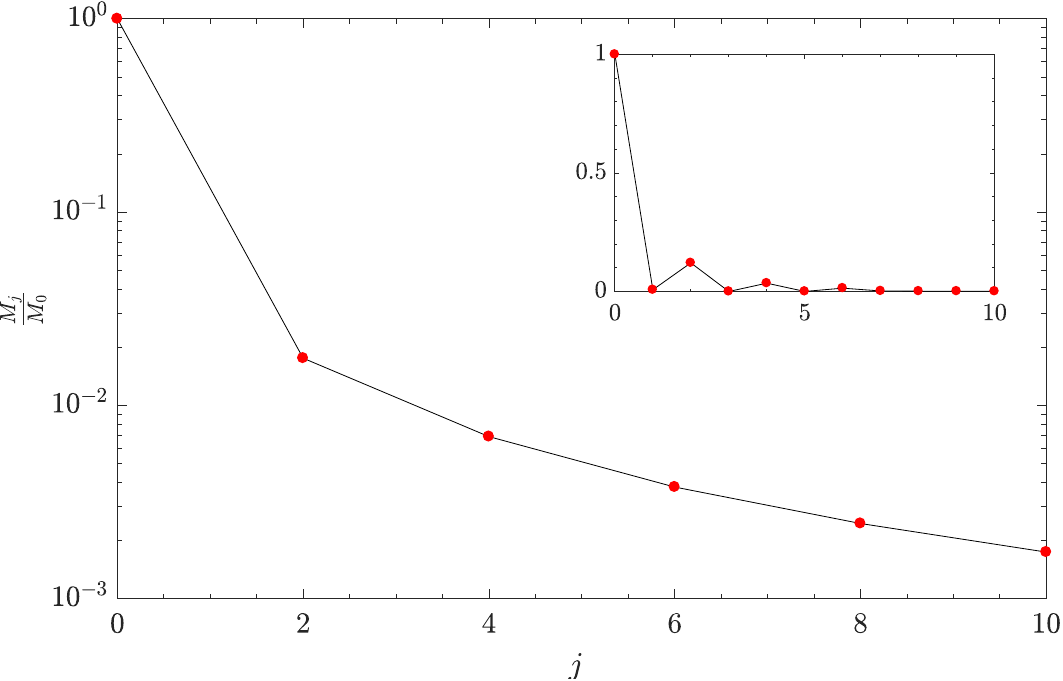}
\caption{Ratio $M_j/M_0$ {\it vs} $j$, for the $x^4$ potential with $N=25$, for the first $10$ even values of $j$. 
The plot is in log--scale. In the inset there is found the same ratio evaluated for the half harmonic oscillator, for every $j$, with $N=25$ in linear--scale.}
\label{fig1}
\end{figure}

Notice that $\rho(x,y)=\rho(-x,-y)$ if the trapping potential $V(x)$ is an even function, therefore in such a case the natural orbitals can be chosen to have definite parity. It turns out that they have the same parity as the single--particle wave functions, \textit{i.e.} $\varphi_j(-x)=(-1)^j\varphi_j(x)$. Then the sum in Eq.~\eqref{n(k=0)} is restricted only to even $j$. For even $j$ the integrals form a decreasing sequence 
\begin{equation}
\label{ordering}
M_0\, >\, M_2 \, > \, M_4 \, > \dots\,,\\
\end{equation}
where the $j=0$ term is typically an order of magnitude greater than $j=2$, which is in turn an order of magnitude greater than $j=4$ term and so on (from hereafter the differences are not that big, but there is still an ordering). In Fig. \ref{fig1} we plot, as an example, the ratios $\frac{M_j}{M_0}$ for the quartic potential and even values of $j=2,4,\dots$. In the inset there is the plot done for the half harmonic oscillator for every $j$ (note the different scales of the two plots). From these figures one can argue that the ordering in (\ref{ordering}) is indeed valid.

\begin{figure}[t]
\includegraphics[width=\columnwidth]{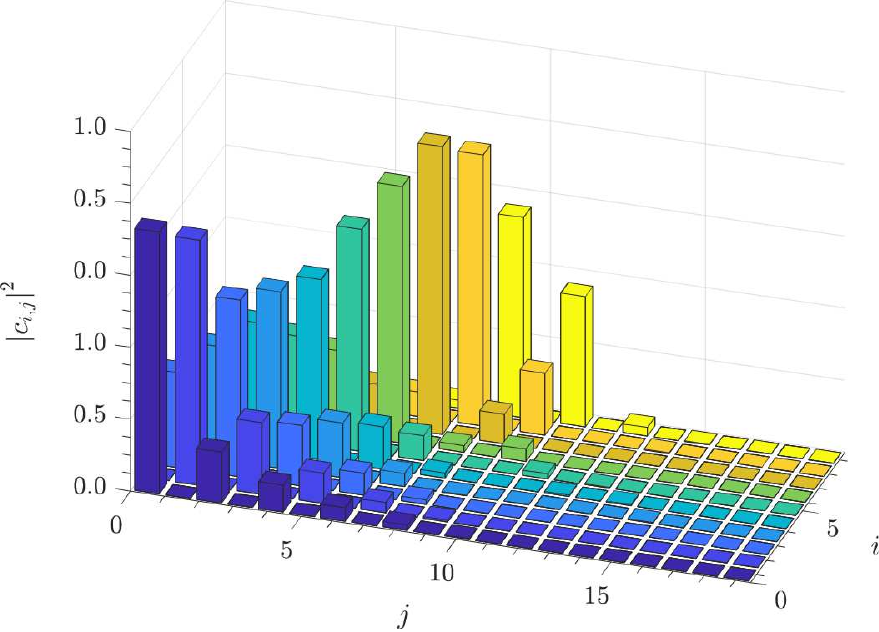}
\caption{First $10$ absolute value squared Fourier coefficients $c_{i,j}$ from (\ref{Fourier_expansion}) weighting the first $20$ eigenfunctions for the 
$x^4$ potential case, \textit{i.e.} $i=0,\dots,9$ and $j=0,\dots,19$ ($N=8$). 
We have explicitly checked that (\ref{Fourier_coeff_norm}) is satisfied up to $1\%$ error.}
\label{fig2}
\end{figure}

To further support this argument, we have also performed an analysis of the coefficients $c_{i,j}$ entering the expansion of the natural orbitals in terms of the single--particle eigenfunctions
\be
\label{Fourier_expansion}
\varphi_i(x)\,=\,\sum_{j} c_{i,j}\,\phi_j(x).
\ee
Since both the sets are orthonormal, the coefficients above have to satisfy 
\be
\label{Fourier_coeff_norm}
\sum_{j} \left | c_{i,j} \right |^2 \,=\,1\,.
\ee
In Fig. \ref{fig2} we plot, as an example, the results for the square of the absolute value of the first Fourier coefficients weighting the first $20$ eigenfunctions for the $V(x) \propto x^4$ potential. 
We conclude that one can write 
\be
\label{n(k=0)_approx}
n_{\text{peak}} \,\approx\,\frac{\lambda_0}{2 \pi} M_0\,.
\ee

\subsection{Scaling laws}
\noindent 
In this Section we aim to determine a relation between the scaling of the largest eigenvalue of the OBDM and the momentum distribution peak. For this purpose, we have studied the behaviour of the natural orbitals $\varphi_j(x)$ for a variety of potentials, including the (even) power--law potentials $V(x) \propto x^{2n}$ and the (non-even) half harmonic  potential $V_{hho}(x)$ defined by $V_{hho}(x) \propto x^2$ for $x>0$ and $V_{hho}(x) = \infty$ for $x\le0$. Due to their normalization, the natural orbitals converge for large values of $N$ to certain functions when the position coordinate $x$ is rescaled by a quantity which depends on (and scale with) $N$. 

\begin{figure}[t]
\includegraphics[width=\columnwidth]{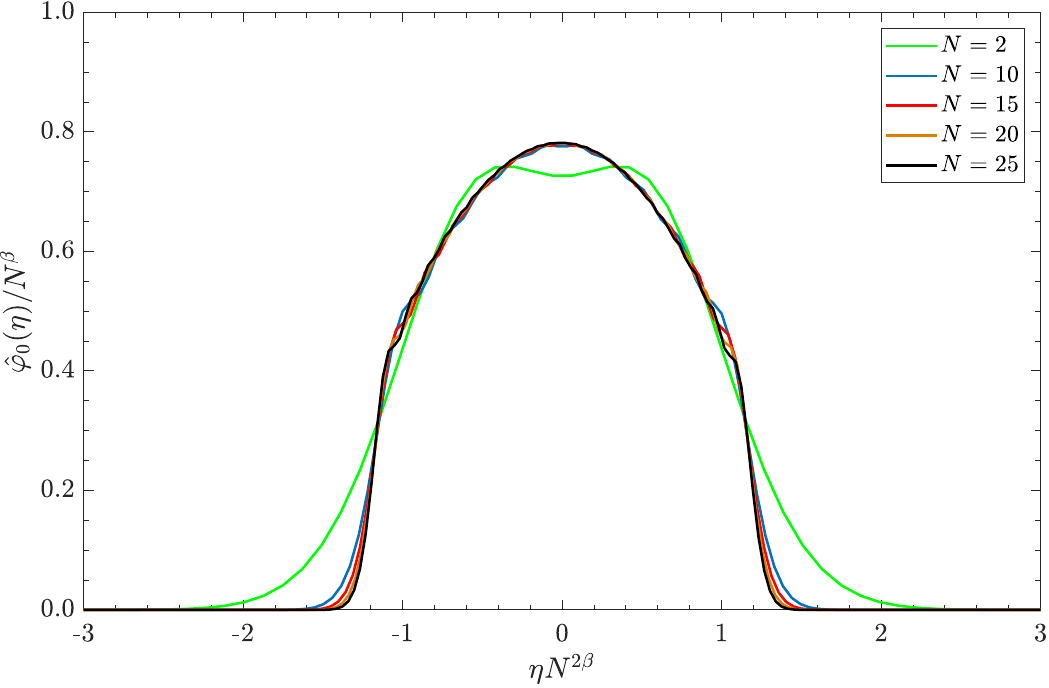}
\includegraphics[width=\columnwidth]{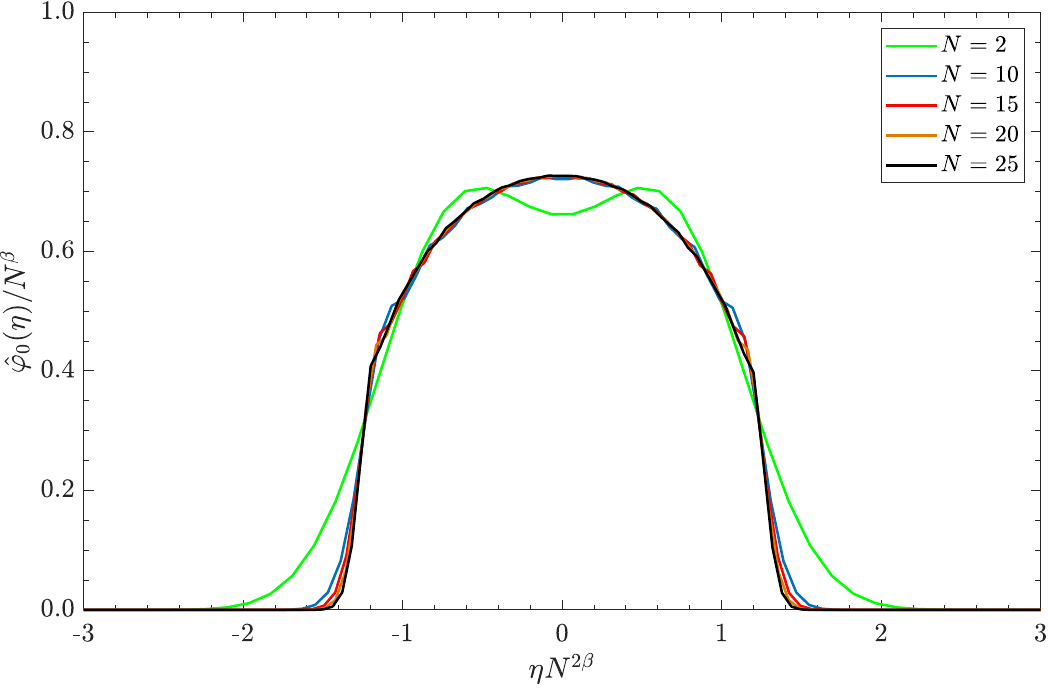}
\caption{$\hat{\varphi}_0(\eta)/N^\beta$ {\it vs} $\eta N^{2 \beta}$ for the harmonic potential ($n=1$) on the top, and the quartic potential ($n=2$) on the bottom. 
From the external part towards the center of both figures, 
we consider $N=2, 10, 15, 20, 25$.}
\label{fig3}
\end{figure}

More precisely, we start by rescaling the position coordinate $x$ in terms of a unit length $\xi$ 
(see Sec.~\ref{sec5} for a definition of $\xi$ in our setup) as 
\be
\label{unit_length}
\eta \, \equiv\, \frac{x}{\xi} \,.
\ee
We then define the dimensionless ground state natural orbital 
$\hat{\varphi}_0(\eta)$ such that
\begin{equation}
\label{norm}
\int\left | \hat{\varphi}_0(\eta) \right |^2 \,d\eta=1.
\end{equation}
It follows that $\hat{\varphi}_0(\eta) \equiv \varphi_0(x)\,\sqrt{\xi}$. We denote by $\beta$ the exponent with which $\hat{\varphi}_0(\eta)$ 
scales with $\hbar$, \textit{i.e.} $\hat{\varphi}_0(\eta)\sim\,\hbar^{-\beta}$; in the semiclassical limit this is equivalent (see Eq.~\eqref{Ndeep}) to  
\be
\label{beta_def}
\hat{\varphi}_0(\eta) \sim \,N^{\beta}\,.
\ee
We have verified that, plotting $\hat{\varphi}_0(\eta)/N^\beta$ as a function of $\eta N^{2 \beta}$, for $N\rightarrow\infty$ the curves converge to a smooth function. 
In Fig. \ref{fig3} we plot $\hat{\varphi}_0(\eta)/N^\beta$ with respect to $\eta N^{2 \beta}$, for the cases of the harmonic  potential (top plot) and quartic potential (bottom plot) for different values
of the particles number. The convergence to a limiting curve for large $N$ is evident from the figures. 

We can also similarly define the dimensionless momentum distribution $n_{\text{peak}}/\xi$ and the exponent $\gamma$ of its scaling with $N$ (or equivalently $\hbar^{-1}$)
\be
\label{n(k=0)scaling_prima}
\frac{n_{\text{peak}}}{\xi}\ \sim N^{\gamma}.
\ee
From Eq.~\eqref{beta_def}, it should be clear that $\int d\eta~\hat{\varphi}_0(\eta)$ must scale as $N^{-\beta}$ for large $N$, in such a way that the normalization condition in Eq.~\eqref{norm} continues to hold. In other words, the support of the function $\hat{\varphi}_0(\eta)$ should scale as $N^{-2\beta}$ (see again Fig.~\ref{fig3}).
From Eqs.~\eqref{n(k=0)_approx} and \eqref{universalresult} we conclude
\be
\label{n(k=0)scaling}
\frac{n_{\text{peak}}}{\xi}\,\sim\,N^{1/2\,-\,2\, \beta}\,.
\ee
In particular, from Eq.~\eqref{n(k=0)scaling} it follows a scaling law among the exponents  $\gamma$ (defined in Eq.~\eqref{n(k=0)scaling_prima}), 
$\beta$ (defined in Eq.~\eqref{beta_def}) and $\mathcal{C}$ (given in Eq.~\eqref{primaryDef})
\be
\label{universal_relation}
\gamma + 2 \beta \,=\,\mathcal{C}\,.
\ee
We will present a numerical check of these results in Sec.~\ref{sec5}, in particular the scaling law (\ref{universal_relation}) for polynomial potentials $V(x) = \Lambda x^{2 n}$ 
with different values of $n$. A prediction for $\gamma$ and $\beta$ for such external trapping potentials will be given at the end of Sec.~\ref{sec5}. For the harmonic potential ($n=1$), from analytical calculations it is already known \cite{Forrester2003,WeiGuLin} that $\mathcal{C}=\frac{1}{2}$, $\beta=-\frac{1}{4}$ and $\gamma=1$, which indeed satisfy both 
Eq.~\eqref{universalresult} and Eq.~\eqref{universal_relation}.

\section{Results for power-law potentials}
\label{sec5}
\subsection{Outline of the numerical technique}
\noindent 
In Sec.~\ref{sec2} we have derived an expression for the OBDM of a TG gas in a generic external potential which leads to Eq.~\eqref{OBDM_final}. In the following, we are going to study the scaling with the particle number of the OBDM maximum eigenvalue $\lambda_0$. For simplicity, we are going to analyse a TG gas at zero temperature trapped by a potential of the form
\be
\label{poly_pot}
V(x)\,=\,\Lambda \, x^{2 n},
\ee 
with $n$ a positive integer and $\Lambda$ a positive coefficient. 

Substituting Eq.~\eqref{poly_pot} into Eq.~\eqref{SP_schro}, one gets a single--particle Hamiltonian with discrete spectrum, and in particular
\be
\left(-\frac{\hbar^2}{2m} \frac{\p^2}{\p x^2} + \Lambda \, x^{2 n}\right)\phi_k(x)\,=\,\eps_k \phi_k(x)\,,
\ee
for $k\,=\,0,\dots,N-1$. It is useful to introduce a length scale $\xi$ through 
\be
\label{def_xi}
\xi=\left(\frac{\hbar^2 b_n}{m\Lambda}\right)^{\frac{1}{2(n+1)}}\,,
\ee
where $b_n$ is a numerical constant that we will fix later. Analogously we define the energy scale 
$\epsilon \,\equiv \, \hbar^2 / (m \xi^2)$ and $\eta=\frac{x}{\xi}$
 [see~Eq.~\eqref{unit_length}] and then rewrite the single--particle Schr\"odinger equation as
\be
\label{SP_schro_AD}
\left[-\frac{1}{2} \frac{\p^2}{\p \eta^2} + b_n \, \eta^{2 n}\right]
\hat{\phi}_k(\eta)\,=\,\frac{\eps_k}{\epsilon} \hat{\phi}_k(\eta)\,. 
\ee
To evaluate the OBDM it is needed to determine the single--particle wave-functions, solutions of Eq.~\eqref{SP_schro_AD}, and substitute their expressions into Eq.~\eqref{OBDM_final}. 
The exact analytical solution of the Schr\"odinger equation (\ref{SP_schro_AD}) is available only for two cases: $n = 1$ and $n = \infty$ that correspond to the harmonic potential and the hard wall, respectively. For intermediate values of $n$, one has to rely either on numerical methods or semiclassical WKB approximation and, as a matter of fact, we have implemented both methods. 

We used the lowest order WKB approximation, WKB$_0$ according to the notation of \cite{BenderOrszag}. One gets then the following estimate for the energy levels of the potential (\ref{poly_pot}) directly from the Bohr--Sommerfeld quantization condition 
\be
\label{WKB_energy}
\eps_k^{wkb} \, = \, \left[\sqrt{\frac{\pi}{2m}}\,\frac{\Gamma\left(\frac{3}{2}+\frac{1}{2n}\right)}{\Gamma\left(1+\frac{1}{2n}\right)}\, \hbar \,\Lambda^{1/2n}\right]^{\frac{2 n}{n +1}} \left(n+\frac{1}{2} \right)^{\frac{2 n}{n +1}}\,,
\ee
where $\Gamma(z)$ is the Euler Gamma function. From Eq.~\eqref{def_xi}, recalling the definition of the length scale $\xi$, we obtain
\be
\label{eps}
\epsilon\,=\,\left[\frac{\hbar}{\sqrt{m}}\,\left(\frac{\Lambda}{b_n} \right)^{\frac{1}{2n}} \right]^{\frac{2n}{n+1}}\,.
\ee
We choose then $b_n$ in Eq.~\eqref{def_xi} in such a way that the energy scale in Eq.~\eqref{eps} matches with the first factor of Eq.~\eqref{WKB_energy}, namely
\be
\label{b(n)}
b_n\,=\,\left[\sqrt{\frac{2}{\pi}}\,\frac{\Gamma\left(1+\frac{1}{2n}\right)}{\Gamma\left(\frac{3}{2}+\frac{1}{2n}\right)}\right]^{2n}\,.
\ee
We have checked (\ref{WKB_energy}) for different values of $n$, comparing the semiclassical results with numerical outcomes obtained with the routine \texttt{Chebfun}~\cite{Chebfun} 
(and also with direct diagonalization of the single--particle Hamiltonian). As one can see from Tab.~\ref{table:energies} and as expected, the WKB formula (Eq.~\eqref{WKB_energy}) approaches the numerical results in the limit of large $k$ (apart of course the harmonic potential case where it is exact). The WKB approximation also provides a form for the 
single--particle wave-functions $\hat{\phi}_k(\eta)$ along the full real line. Near the turning points of the potential, one has to use a standard  Airy function approximation.  
\begin{table}[t]
\begin{tabular}{c|c|c||c|c||c|c||c|c||c|c|}
\cline{2-7}
&\multicolumn{2}{c||}{$n=2$} & \multicolumn{2}{c||}{$n=3$} & \multicolumn{2}{c||}{$n=4$}  \\ \hline
\multicolumn{1}{|c|}{$k$}  & $\eps_k^{wkb}$ & $\eps_k^{cheb}$ & $\eps_k^{wkb}$ & $\eps_k^{cheb}$ & $\eps_k^{wkb}$ & $\eps_k^{cheb}$  \\ \hline
\multicolumn{1}{|c|}{$0$}  &   $\,0.397\,$        &     $\,0.485\,$       &     $\,0.353\,$       &  $\,0.505\,$         &    $\,0.330\,$        &  $\,0.530\,$                  \\ \hline
\multicolumn{1}{|c|}{$1$}    & $1.717$          &   $1.739$         &     $1.837$       &   $1.915$        &    $1.913$        &      $2.059$           \\ \hline
\multicolumn{1}{|c|}{$2$}  &  $3.393$         &    $3.412$        &    $3.953$         &  $4.006$        &   $4.332$         &     $4.435$              \\ \hline
\multicolumn{1}{|c|}{$3$}  &   $5.314$        &    $5.329$        &       $6.548$      &   $6.593$       &   $7.421$         &     $7.509$          \\ \hline
\multicolumn{1}{|c|}{$4$}  &    $7.429$       &   $7.442$         &     $9.546$       &  $9.586$         &    $11.09$        &   $11.17$               \\ \hline
\multicolumn{1}{|c|}{$5$}  &  $9.708$         &    $9.720$        &      $12.89$       &    $12.93$      &    $15.30$        &  $15.37$                \\ \hline
\multicolumn{1}{|c|}{$6$} &   $12.13$        &    $12.14$        &   $16.57$         &  $16.61$         &   $19.98$         &    $20.05$             \\ \hline
\multicolumn{1}{|c|}{$7$} &   $14.68$        &    $14.69$        &   $20.54$         &    $20.57$       &   $25.12$         &     $25.19$            \\ \hline
\multicolumn{1}{|c|}{$8$} &     $17.35$      &    $17.35$        &    $24.78$        &     $24.81$      &    $30.69$        &     $30.75$             \\ \hline
\multicolumn{1}{|c|}{$9$} &     $20.12$      &   $20.13$         &    $29.28$        &     $29.31$      &      $36.67$      &   $36.73$           \\ \hline
\multicolumn{1}{|c|}{$10$} & $22.99$          &    $23.00$        &    $34.02$        &   $34.05$        &    $43.04$        &    $43.10$          \\ \hline

\end{tabular}
\caption{Semiclassical energy levels obtained from (\ref{WKB_energy}) and the 
corresponding numerical results for $\eps_k$ for $n = 2, 3, 4 $ and 
$V(\eta)=b_n\,\eta^{2n}$, with $b_n$ fixed by (\ref{b(n)}). Energies 
are in units of $\epsilon$.}
\label{table:energies}
\end{table}

After determining the single--particle wave-functions either numerically or within the WKB approximation, we have generated the OBDM (\ref{OBDM_final}) for the potentials in Eq.~\eqref{poly_pot}. For $n=1$ and $n=\infty$ exact form of the wave functions are of course available and the task simplifies. Finally, we are left with the eigenvalue problem
\be
\label{diagonalization_problem}
\int \rho(x,y)\, \varphi_j(y) \,dy\,=\,\lambda_j\, \varphi_j(x)\,,
\ee 
that we have solved by discretizing the integral; for finite $n$, for instance, one can employ a Gauss--Hermite quadrature~\cite{Delves} (see also the Appendix \ref{app:A} for  more 
details). To be sure that the method works accurately for different potentials and particles number, we have verified whether the results converge increasing the number of 
nodes (points) of the quadrature. 

We are interested in the study of deviations from ODLRO and  therefore we focus our attention on the behaviour of $\lambda_0$ for different number of particles in the system. To characterise and quantify these deviations in the TG gas, we have fitted the large $N$ asymptotic of the maximum eigenvalue of the OBDM with a power--law \cite{Forrester2003} 
\be
\label{fitting_powerlaw}
\lambda_0\,=\,\mathcal{A}+\mathcal{B}\,N^{\mathcal{C}}+\frac{\mathcal{D}}{N^{\mathcal{E}}}\,,
\ee
where in principle all the parameters $\mathcal{A},\dots,\mathcal{E}$ are potential-dependent (\textit{i.e.} $n$-dependent). Since the number of particles $N$ typically goes from $2$ to $25$, sub-leading finite-size corrections are taken into account by the exponent $\mathcal{E}$ (and the pref-actor $\mathcal{D}$) in Eq.~\eqref{fitting_powerlaw}.

As discussed in the Introduction, it is possible to define two different scalings of $\lambda_0$ with respect to the particle number. In the first case (case $(b)$), we could fix the external potential and increase $N$. In the second case (case $(a)$) we could fix instead the density of particles in the trap and vary $N$ and $\Lambda$ accordingly. For example, for the harmonic potential we can write 
$\Lambda \, = \, \frac{1}{2} \,m\,\omega^2$, and, using the length scale $\xi\, \equiv \,\sqrt{\hbar / m \omega}$, we can define the average density $\rho\,=\,N /\sqrt{\hbar / m \omega}$. 
We are going to approach the problem in both ways.

A power--law scaling similar to Eq.~\eqref{fitting_powerlaw} can be also argued for the dimensionless momentum distribution peak
\be
\label{FGHI_def}
\frac{n_{\text{peak}}}{\xi} \,=\,\mathcal{F}+ \mathcal{G}\,N^{\gamma}+\frac{\mathcal{H}}{N^{\mathcal{I}}} \,.
\ee
To obtain $\beta$ defined in (\ref{beta_def}), we proceed in the following way. First we evaluate $\hat{\varphi}_0(\eta)$ for two different values of the particle number, $N_1$ and $N_2$. To have an estimate of $\beta$, we impose that 
\be
\hat{\varphi}_0^{(N_1)}(\eta) \, N_1^{-\beta} \,=\,\hat{\varphi}_0^{(N_2)}(\eta) \, N_2^{-\beta}\,,
\ee
near the origin, from which it follows that
\be
\label{beta_metodo}
\beta\,=\,\frac{\ln\left[\hat{\varphi}_0^{(N_2)}(\eta)\, \big/\, \hat{\varphi}_0^{(N_1)}(\eta)\right]}{\ln\left(N_2\big/{N_1}\right)}\,.
\ee
Once the value of $\beta$ is found, we have checked that the scaled natural orbitals, \textit{i.e.}  $\hat{\varphi}_0(\eta) \, N^{-\beta}$, converge by increasing $N$.

\subsection{Semiclassical (CFT) determination of the pre-factor $\mathcal{B}$ in Eq.~\eqref{primaryDef}}
\label{sub:pred_B}
\noindent
Let's now show how it is possible to use the asymptotic form in Eq.~\eqref{Dubailformula} for the OBDM to extract directly the potential-dependent coefficient $\mathcal{B}$ in Eq.~\eqref{fitting_powerlaw} for $N\rightarrow\infty$ [see~Eq.~\eqref{universalresult}]. Once again we focus on power-law potentials given in Eq.~\eqref{poly_pot}. In the semiclassical limit defined in Eq.~\eqref{slimit}, the following dimensionful quantities $(\mu,\Lambda,m)$ do not scale with $\hbar$, and we replace them with $(R,\tilde{L},m)$ where $R\equiv\left(\frac{\mu}{\Lambda}\right)^{1/2n}$ is a length scale and $\tilde{L}$ the time scale in Eq.~\eqref{LDubail}. For the power-law potentials
\begin{equation}
\tilde{L}=\Xi R \sqrt{\frac{m}{2\mu}},
\end{equation} 
where $\Xi$ a numerical constant given by
\begin{equation}
\Xi=\int_{-1}^{1}\frac{du}{\sqrt{1-u^{2n}}}=2\sqrt{\pi}
\frac{\Gamma\left(\frac{2n+1}{2n}\right)}{\Gamma\left(\frac{n+1}{2n}\right)}.
\end{equation}
In the semiclassical approximation $\hbar$ can be replaced by $N$ according to Eq.~\eqref{Ndeep} which in our case reads
\begin{equation}
\label{ndep_pot}
N\hbar=\frac{mR^2}{\tilde{L}}\left(\frac{\alpha\Xi}{\pi}\right).
\end{equation}
In Eq.~\eqref{ndep_pot}, $\alpha$ is another numerical constant given by
\begin{equation}
\alpha=\int_{-1}^{1}du~\sqrt{1-u^{2n}}=\sqrt{\pi}
\frac{\Gamma\left(\frac{2n+1}{2n}\right)}{\Gamma\left(\frac{3n+1}{2n}\right)}.
\end{equation}
The OBDM in Eq.~\eqref{Dubailformula} is expressed in terms of a variable $\tilde{x}(x)\equiv\tilde{L}F(x/R)$.  Again for the potentials in Eq.~\eqref{poly_pot}, $F$  is given by
\begin{equation}
\label{F}
F(\eta)\,=\, \frac{1}{\Xi}\int_{-1}^{\eta}du~\frac{1}{\sqrt{1-u^{2n}}}.
\end{equation}
However, it is actually more convenient to introduce $G(\eta)$ in such a way that $F(\eta)\equiv\frac{1}{2}+G(\eta)$ and it turns out
\begin{equation}
\label{G}
G(\eta)\,=\,\text{sign}(\eta)\frac{B_{\eta^{2n}}\left(\frac{1}{2n},\frac{1}{2}\right)}{2n\Xi},
\end{equation}
where the function $B_z(a,b)$ is the incomplete Beta function  \cite{AS}. 
The function $G(\eta)$ simplifies in the limit $n=1$ (harmonic oscillator) where we have $G(\eta)|_{n=1}=\frac{1}{\pi}\arcsin(\eta)$ and also in the limit $n\rightarrow\infty$ (hard wall) where  $G(\eta)|_{n=\infty}=\frac{1}{2}\eta$. Taking into account all of this, we can rewrite the eigenvalue equation \eqref{eigen_eq} for the semiclassical OBDM as
\begin{equation}
\label{semic_eq}
\sqrt{N}|C|^2\sqrt{\frac{\pi}{\alpha\Xi}}\int_{-1}^{1}d\eta'~K(\eta,\eta')\varphi_j(\eta')=\lambda_j\varphi_j(\eta)\,,
\end{equation}
where $K(\eta,\eta')$ is the kernel
\begin{equation}
\label{ker}
K(\eta,\eta')\,=\, \frac{\left | 1-\sin^2(\pi G(\eta)) \right|^{\frac{1}{8}}\ \left | 1-\sin^2(\pi G(\eta'))\right |^{\frac{1}{8}}}{\left | \sin(\pi G(\eta))-\sin(\pi G(\eta')) \right  |^{\frac{1}{2}}}\,.
\end{equation}
As already anticipated, the existence of the limit~\eqref{slimit} requires $\lambda_j\sim \mathcal{B}_j\sqrt{N}$, \textit{i.e.} $\mathcal{C}=1/2$ [see~Eq.~\eqref{fitting_powerlaw}]. The numerical pre-factors $\mathcal{B}_j$ can be calculated from the knowledge of the eigenvalues $\bar{\lambda}_j$ of Eq.~\eqref{ker}. Indeed from (\ref{semic_eq}) one has
\begin{equation}
\label{bfact}
\mathcal{B}_j\equiv|C|^2\sqrt{\frac{\pi}{\alpha\Xi}}\bar{\lambda}_j.
\end{equation}
In the following we only focus on the scaling of largest eigenvalue $\lambda_0$ and then define $\mathcal{B}\equiv\mathcal{B}_0$, consistently with Eq.~\eqref{primaryDef}. 

The largest eigenvalue of the kernel in Eq.~\eqref{ker} can be obtained with a numerical procedure similar to the one outlined at the end of the previous Section. 
Notice that the kernel in Eq.~\eqref{ker} is singular for $\eta=\eta'$ and its diagonal elements have to be regularized with a cut-off $\delta$. The physical origin of the cut-off can be traced back to the validity of Eq.~\eqref{Dubailformula} up to distances $|x-y|\simeq\rho_{\text{max}}^{-1}$. In the dimensionless variable $\eta$, therefore the cut-off is
$\delta\simeq\frac{\rho_{\text{max}}^{-1}}{R}\ll 1$. This condition, determining the validity of the CFT approach, already appears in \cite{BrunDubail}. From this perspective the semiclassical limit $\hbar\rightarrow 0$ in Eq.~\eqref{slimit} it is actually a convenient way to take the continuum limit to a field theory. Such a field theory describes the gas density fluctuations on intermediate length scales much larger than $\rho_{\text{max}}^{-1}$ and much smaller than the effective length $R$ of the system~\cite{Calabrese2017, BrunDubail, Brun18}. A non-trivial consequence is that the  two procedures $(a)$ and $(b)$ to implement the large-$N$ limit should reproduce the same results. Indeed fixing the external potential and varying the density is equivalent to consider $\rho_{\max}^{-1}\rightarrow 0$ while keeping $R$ fixed;  on the other hand fixing the density and varying the potential corresponds to $R\rightarrow\infty$ while keeping $\rho_{\max}$ fixed. In both cases $\delta\ll 1$ and the gas is described by a CFT. 

\begin{figure}[t]
\includegraphics[width=\columnwidth]{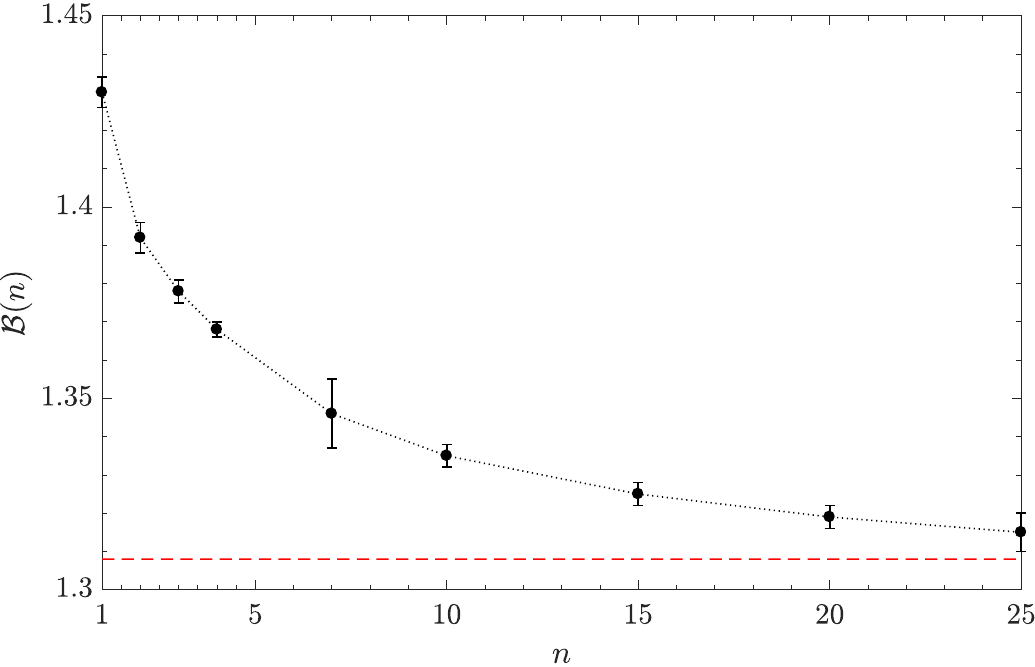}
\caption{$\mathcal{B}(n)$ {\it vs} $n$ obtained 
from solving Eq.~\eqref{semic_eq} with kernel given by Eq.~\eqref{ker}. $\mathcal{B}$ monotonically decreases from $1.430(4)$ for $n=1$ to $1.308(3)$ for $n=\infty$ which is represented by the red dashed line. The black dotted line 
is a guide for the eye.}
\label{fig4}
\end{figure} 

Numerical estimations for $\mathcal{B}$ in Eq.~\eqref{bfact} obtained with a Gauss-Legendre quadrature up to $Z=20000$ nodes are given in Tab.~\ref{table:checkB} and Fig. \ref{fig4}. $Z$ is 
the number of points of the grid in which the interval $[-1,1]$ is divided. The error is estimated by extrapolating the value of $\mathcal{B}$ in the limit $\delta \to 0$ by increasing 
$Z$. Then the obtained values for varying $Z$ are fitted with a function of the form $\mathcal{B}+{\mathcal M}/Z^{\zeta}$. We observe that Refs.~\cite{Forrester2003, ForresterJournal2003} also provide a numerical evaluation of $\mathcal{B}$ for the harmonic potential ($n=1$) and the hard wall ($n=\infty$). Our results fully confirm and generalize these predictions.

The CFT predictions for $\mathcal{B}$ are compared in Tab.~\ref{table:checkB} with $\mathcal{B}_{\text{fit}}$, which is the value of $\mathcal{B}$ obtained from the fit (\ref{fitting_powerlaw}) using the numerical results for the OBDM $\rho(x,y)$ directly computed. The large-$N$ limit is implemented here fixing the potential and varying the density (case $(b)$). Notice that in doing the fit one could either fix $\mathcal{C}$ to the value $1/2$ or re-fit as well $\mathcal{C}$ according to (\ref{fitting_powerlaw}). Since the value $\mathcal{C}=1/2$ has been independently established and checked, we present our results for $\mathcal{B}_{\text{fit}}$ with the former procedure. When instead $\mathcal{C}$ 
is re-fitted, substantial agreement for $\mathcal{B}_{\text{fit}}$ is found, except for $n \sim 2 - 4$ where we obtained a discrepancy of order $1\%$. To check which procedure is better, we performed both with $N$ up to $30$, and then we compared their predictions with the value for $\lambda_0$ obtained for $N=35$ directly from the numerical diagonalization of the ODBM. We found that the procedure in which $\mathcal{C}$ is fixed gives slightly better results. Finally we also verified that the scaling of the largest eigenvalue obtained fixing the density in the trap and varying the potential (case $(a)$) is also consistent with the CFT predictions; see Table~\ref{table:fixeddensity}, for fits without fixing $\mathcal{C}=1/2$ and Table~\ref{table:fixeddensityfixedC} for fits with $\mathcal{C}=1/2$. Compare in particular the results in Table~\ref{table:fixeddensityfixedC} with those collected in Table~\ref{table:FittingResultsLambda}.

In conclusion, the agreement between the predictions obtained from the CFT formula~\eqref{Dubailformula} and the numerical values for $\mathcal{B}_{\text{fit}}$ is very satisfactory.

\subsection{Numerical results}
\label{sub:num_results}
\noindent
Let's now describe the outcome of the numerical analysis, based on Eq.~\eqref{fitting_powerlaw} and Eq.~\eqref{FGHI_def}, for the large-$N$ behaviour of $\lambda_0$, $n_{\text{peak}}/\xi$ and $\hat{\varphi}_0(\eta)$. Such a study strongly corroborates the validity of Eq.~\eqref{universalresult} and Eq.~\eqref{universal_relation}. To obtain the results in Tab.~\ref{table:FittingResultsLambda}, Tab.~\ref{table:FittingResultsMomDistr} and Tab~\ref{table:finalresults}, we have varied the density of particles in the system (by increasing $N$ typically up to $25 - 30$) for different power-law potentials in Eq.~\eqref{poly_pot}. In particular
\begin{itemize}

\item In Tab.~\ref{table:FittingResultsLambda} are collected the results obtained for the parameters of the scaling of $\lambda_0$ according to Eq.~\eqref{fitting_powerlaw}. 

\item In Tab.~\ref{table:FittingResultsMomDistr} we report the results for the dimensionless momentum distribution peak according to Eq.~\eqref{FGHI_def}.

\item  In Tab.~\ref{table:finalresults} we summarise the values of $\mathcal{C}$, $\beta$ and $\gamma$ obtained as a function of $n$.

\end{itemize}

\begin{table}[t]
\begin{center}
\centering
\begin{tabular}{|c|c|c|}
\hline
$ n$ & $\mathcal{B}_{\text{fit}}$ & $\mathcal{B}$ \\ \hline
$1$ & $\,\,1.4304(2)\,\,$ & $\,\,1.430(4)\,\,$ \\ \hline
$2$ & $1.400(4)$ & $1.392(4)$ \\ \hline
$3$ & $1.380(4)$ & $1.378(3)$ \\ \hline
$4$ & $1.372(5)$ & $1.368(2)$ \\ \hline
$\infty$ & $1.31(1)$ & $1.308(3)$ \\ \hline 
\end{tabular}
\caption{Estimation of the pre-factor $\mathcal{B}$ in Eq.~\eqref{primaryDef}. The values $\mathcal{B}_{\text{fit}}$ are obtained fitting the finite-$N$ results for the largest eigenvalue of Eq.~\eqref{OBDM_final} with Eq.~\eqref{fitting_powerlaw}. The fit for $\lambda_0$ is done fixing the potential and varying the density by increasing the number of particles.
The numerical error in the last digit is reported in brackets. In the second column are given the values of $\mathcal{B}$ obtained from Eq.~\eqref{bfact}, after determining
numerically the largest eigenvalue $\bar{\lambda}_0$ of the kernel in Eq.~\eqref{ker}. }
\label{table:checkB}
\end{center}
\end{table}

\begin{table}[t]
\begin{center}
\centering
\begin{tabular}{|c|c|c|c|c|}
\hline
$n$ & $\mathcal{A}_{\text{fit}}$  & $\mathcal{B}_{\text{fit}}$ & $\mathcal{D}_{\text{fit}}$ & $\mathcal{E}_{\text{fit}}$  \\ \hline
$1$      &       $-0.554(2)$   &   $\,\,1.4304(2)\,\,$         &     $0.122(1)$     &   $\,\,0.60(1)\,\,$    \\ \hline
$2$      &       $-0.55(4)$     &   $1.400(4)$           &     $0.141(8)$     &   $\,0.79(6)\,$        \\ \hline
$3$      &       $\,\,-0.53(3)\,\,$      &   $\,\,1.380(4)\,\,$           &     $\,\,0.16(2)\,\,$        &   $1.1(5)$            \\ \hline
$4$     &       $-0.56(3)$      &   $1.372(5)$             &     $0.20(1)$          &   $0.9(3)$         \\ \hline
$\infty$ &       $-0.6(1)$     &   $1.31(1)$         &     $0.31(9)$          &   $0.3(1)$             \\ \hline
\end{tabular}
\caption{Results for the parameters entering Eq.~\eqref{fitting_powerlaw} for different values of $n$ by fixing the trapping potential and varying the density at the center of the trap. 
The numerical error in the last digit is reported in brackets. 
}
\label{table:FittingResultsLambda}
\end{center}
\end{table}

\begin{table}[t]
\begin{center}
\centering
\begin{tabular}{|c|c|c|c|c|}
\hline
$n$ & $\mathcal{F}_{\text{fit}}$  & $\mathcal{G}_{\text{fit}}$ & $\mathcal{H}_{\text{fit}}$ & $\mathcal{I}_{\text{fit}}$  \\ \hline
$1$      &      $0.002(2)$         &  $0.561(6)$       &          &                 \\ \hline
$2$      &         $\,-0.046(8)\,$      &   $\,\,0.5001(3)\,\,$      &   $\,\,0.025(6)\,\,$      &      $\,0.6(3)\,$             \\ \hline
$3$      &         $-0.15(8)$      &  $0.491(2)$       &    $0.10(8)$      &      $0.3(3)$            \\ \hline
$4$      &           $-0.21(1)$    &    $0.500(5)$     &     $0.015(5)$     &     $0.41(3)$             \\ \hline
$\infty$ &         $-0.752(3)$      &    $0.1994(2)$     &   $\,\,0.0048(1)\,\,$       &       $\,\,1.09(5)\,\,$            \\ \hline
\end{tabular}
\caption{Results for the parameters in the dimensionless 
momentum distribution peak (\ref{FGHI_def}) for different values of $n$. 
Note that for $n = 1$ the correction term $\propto 1/N$ in the fitting is not necessary. }
\label{table:FittingResultsMomDistr}
\end{center}
\end{table}

\begin{table}[b]
\begin{center}
\centering
    \begin{tabular}{c|cccc||cc}
    $n$ & $\mathcal{C}_{\text{fit}}$ & $\mathcal{C}^{wkb}$ &$\beta_{\text{fit}}$ & $\gamma_{\text{fit}}$ &  $\beta$ & $\gamma$ \\ \hline
    $1$   & $\,\,0.500(2)\,\,$ & $\,\,0.496(8)\,\,$ & $\,\,-0.25(1)\,\,$    & $\,\,1.02(4)\,\,$  & $\,\,-\frac{1}{4}\,\,$    & $\,\,1\,\,$     \\
    $2$    & $0.501(1)$ & $\,\,0.54(3)\,\,$ & $-0.16(1)$    & $0.85(2)$   & $-\frac{1}{6}$    & $\frac{5}{6}$     \\
    $3$  & $0.501(2)$ & $\,\,0.54(7)\,\,$ & $-0.12(2)$    & $0.76(1)$     & $-\frac{1}{8}$    & $\frac{3}{4}$     \\
    $4$   & $0.500(3)$ & $\,\,0.54(9)\,\,$ & $-0.10(1)$    & $0.70(1)$     & $-\frac{1}{10}$    & $\frac{7}{10} $    \\
 $\infty$ & $0.500(1)$ & & $0.00(1)$    & $0.502(2)$  & $0$    & $\frac{1}{2}$     \\
    \end{tabular}
\caption{From the second to the fifth columns are gathered numerical results 
for the parameters ruling the scaling with $N$ of $\lambda_0$, $\lambda_0^{wkb}$, $\hat{\varphi}_0(\eta)$ and $n_{\text{peak}}/\xi$. 
The value for $\mathcal{C}$, within the numerical precision, appears to be independent on the potential and equal to $1/2$.
Last two columns: Exact values for $\beta$ and $\gamma$ coming from Eq.~\eqref{guessing_betan} and 
Eq.~\eqref{guessing_gamman} respectively.}
\label{table:finalresults}
\end{center}
\end{table}

\begin{figure}[t]
\includegraphics[width=\columnwidth]{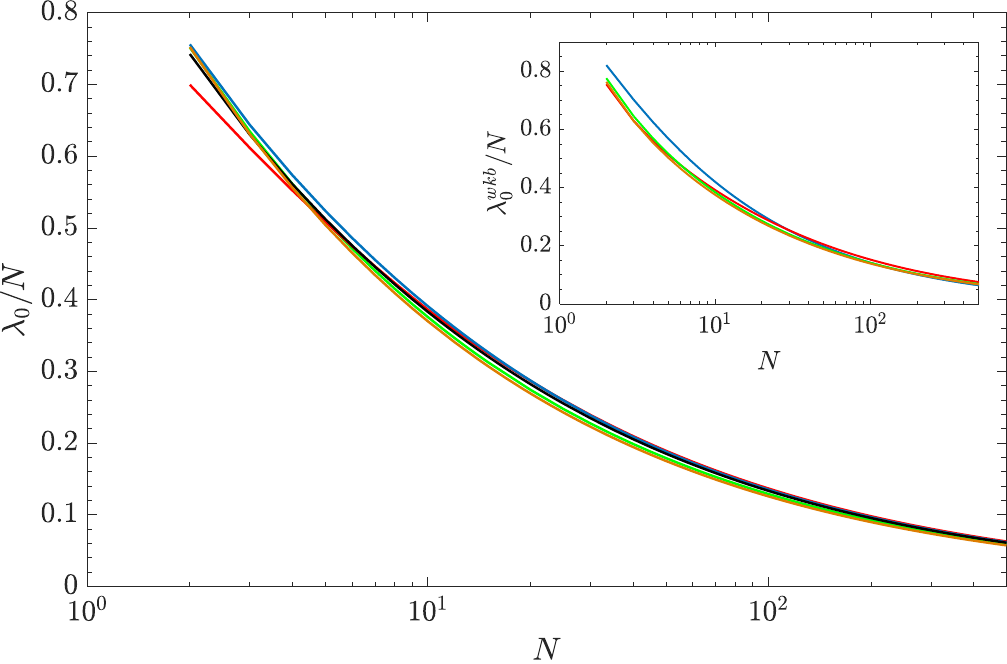}
\caption{$\lambda_0$ {\it vs} $N$ obtained 
for $n = 1$ (in blue), $n = 2$ (in red), $n = 3$ (in green), $n = 4$ (in orange) and $n = \infty$ (in black), up to $500$ particles. In the inset is shown the behaviour of $\lambda_0^{wkb}$ vs $N$ obtained with the WKB approximation.}
\label{fig5}
\end{figure}

Our findings for $\lambda_0$ as a function of $N$ using Eq.~\eqref{fitting_powerlaw} are plotted for different values of $n$ in Fig. \ref{fig5}. The inset of Fig.~\ref{fig5} shows the WKB approximation results, where it is called $\lambda_0^{wkb}$. 
The exponent $\mathcal{C}$ is approaching $1/2$ within the numerical error.

By studying the system by fixing the density and varying $\Lambda$, we have collected the data reported in Table \ref{table:fixeddensity} for different values of $n$ of the polynomial trapping potential. From these results is evident that $\mathcal{C}=1/2$ is also found in this case. For this reason we then fitted the data via Eq.~\eqref{fitting_powerlaw} with $\mathcal{C}$ fixed to the value $1/2$ and we obtained the results reported in Table \ref{table:fixeddensityfixedC}. Consistently with the discussion in Sec.~\ref{sub:pred_B}, the values estimated for the parameter $\mathcal{B}$ (actually for all the fit parameters) are consistent within the error bar with the ones in Table~\ref{table:FittingResultsLambda}.


\begin{table}[t]
\begin{center}
\centering
\begin{tabular}{|c|c|c|c|c|c|}
\hline
$n$ & $\mathcal{A}_{\text{fit}}$  & $\mathcal{B}_{\text{fit}}$ & $\mathcal{C}_{\text{fit}}$ & $\mathcal{D}_{\text{fit}}$ & $\mathcal{E}_{\text{fit}}$  \\ \hline
$1$      &       $\,\,-0.58(2)\,\,$   &   $\,\,1.45(3)\,\,$     & $\,\,0.4995(8)\,\,$    &     $\,\,0.36(5)\,\,$     &   $\,\,2.00(1)\,\,$    \\ \hline
$2$      &       $-0.56(3) $     &   $1.418(2) $       &  $ 0.498(5)$       &     $0.16(4) $     &   $1.2(3) $        \\ \hline
$3$      &       $-0.56(2) $      &   $1.391(1) $      &    $0.500(3)  $      &     $ 0.18(2)$        &   $ 1.0(1)$            \\ \hline
$4$     &       $-0.52(7) $      &   $1.34(4) $     & $ 0.503(6)$        &     $0.21(1) $          &   $0.30(5) $         \\ \hline
\end{tabular}
\caption{Results for the parameters entering Eq.~\eqref{fitting_powerlaw} for different values of $n$ by fixing the density at the center of the traps and varying $N$ and $\Lambda$ accordingly. 
The numerical error in the last digit is reported in brackets. 
}
\label{table:fixeddensity}
\end{center}
\end{table}

\begin{table}[t]
\begin{center}
\centering
\begin{tabular}{|c|c|c|c|c|}
\hline
$n$ & $\mathcal{A}_{\text{fit}}$  & $\mathcal{B}_{\text{fit}}$ & $\mathcal{D}_{\text{fit}}$ & $\mathcal{E}_{\text{fit}}$  \\ \hline
$1$      &       $\,\,-0.56(3)\,\,$   &   $\,\,1.432(4)\,\,$      &     $\,\,0.13(3)\,\,$     &   $\,\,0.57(2)\,\,$    \\ \hline
$2$      &       $-0.55(2) $     &   $1.407(4) $             &     $0.15(3) $     &   $1.0(2) $        \\ \hline
$3$      &       $-0.56(2) $      &   $1.391(1) $          &     $ 0.18(2)$        &   $ 1.0(1)$            \\ \hline
$4$     &       $-0.56(1) $      &   $1.38(2) $             &     $0.18(4) $          &   $0.8(1) $         \\ \hline
\end{tabular}
\caption{Results for the parameters entering Eq.~\eqref{fitting_powerlaw} with $\mathcal{C}$ fixed to $1/2$, for different values of $n$ by fixing the density at the center of the traps and varying $N$ and $\Lambda$ accordingly. 
The numerical error in the last digit is reported in brackets. 
}
\label{table:fixeddensityfixedC}
\end{center}
\end{table}

We have also done calculations for a potential $V_{hho}(x)=\Lambda x^{2}$ for $x>0$ and zero otherwise, \textit{i.e.} half of the harmonic potential: By varying the density in the system we get
\begin{align}
&\mathcal{A}_{\text{fit}}=-0.37(1); \,\,\, \mathcal{B}_{\text{fit}}=1.267(3); \,\,\, \mathcal{C}_{\text{fit}}=0.51(2);
\nonumber\\
&\mathcal{D}_\text{fit}=0.103(1);\,\,\, \mathcal{E}_{\text{fit}}=0.58(1) \,.
\end{align}

\subsection{The hard wall potential}
\label{sub:hard_wall}
\noindent
The case of $n=\infty$ is analogous to impose Dirichlet boundary conditions (DBC) and the situation is slightly different than the previous cases. In this case, one has to evaluate 
\be
\label{diagonalization_finiteL}
\int_{0}^{L} \rho(x,y)\, \varphi_j(y) \,dy\,=\,\lambda_j\, \varphi_j(x)\,,
\ee 
so that the Gauss--Hermite quadrature cannot be applied any more. We rather used the Gauss--Legendre quadrature, see Appendix \ref{app:A}. The same will happen with periodic boundary conditions (PBC) and Neumann boundary conditions (NBC). In these cases Vandermonde determinant formulas can be used to get closed expressions for the OBDM \cite{ForresterCommunications2003}, which are easier to handle numerically for large number of particles [still the formula (\ref{OBDM_final}) can be used]. Therefore one just needs 
to construct the entire OBDM varying $\theta=\pi x / L$ and $\sigma=\pi y / L$ from $0$ to $\pi$ and directly diagonalize the finite dimensional matrix after the discretization. 
With these three different boundary conditions we got the following results:
\begin{itemize}
\item PBC: Using the results in \cite{CMT2018}, we can compute the eigenvalues of the OBDM for a TG gas in a circular geometry up to $10^3$ particles. 
In this case the best fitting law is the one not having the correction term $\propto 1/N$ in (\ref{fitting_powerlaw}), since we work with very high number of particles. In this case we have
\begin{align}
\label{PBC_fits}
&\mathcal{A}_{\text{fit}}=-0.597(1); \,\,\, \mathcal{B}_{\text{fit}}=1.4741(1);\nonumber\\
&\mathcal{C}_{\text{fit}}=0.5000(1) \,.
\end{align}
Fixing $\mathcal{C} =1/2$ in the fitting procedure, one gets
\begin{align}
\label{PBC_fits}
&\mathcal{A}_{\text{fit}}=-0.602(1); \,\,\, \mathcal{B}_{\text{fit}}=1.475(1)\,.
\end{align}

\item DBC ($n=\infty$): With the hard wall potential with $b_\infty=0$ from (\ref{b(n)}), we have considered values of $N$ up to $N=150$. The corresponding results for $\lambda_0$ are 
presented in Table \ref{table:FittingResultsLambda} and \ref{table:finalresults}, and those for the momentum distribution in Table \ref{table:FittingResultsMomDistr}. For the natural orbitals 
we found that they are independent of $N$, \textit{i.e.} $\beta\,=\,0$. 
\item NBC: In this case we computed the OBDM and its eigenvalues up to $100$ particles. 
Fitting via Eq.~\eqref{fitting_powerlaw}, we have 
\begin{align}
&\mathcal{A}_{\text{fit}}=-0.71(1); \,\,\, \mathcal{B}_{\text{fit}}=1.340(4); \,\,\, \mathcal{C}_{\text{fit}}=0.497(5)\nonumber\\
&\mathcal{D}_\text{fit}=0.482(7); \,\,\, \mathcal{E}_\text{fit}=1.33(5)\,.
\end{align}
Fixing $\mathcal{C} = 1/2$ during the fitting procedure, we get
\begin{align}
&\mathcal{A}_{\text{fit}}=-0.68(5); \,\,\, \mathcal{B}_{\text{fit}}=1.32(1); \nonumber\\
&\mathcal{D}_\text{fit}=0.49(7); \,\,\, \mathcal{E}_\text{fit}=1.6(2)\,.
\end{align}

\end{itemize}

\subsection{Analytical predictions for $\beta$ and $\gamma$}
\label{sub:analyt}
\noindent
All  previous results are compatible (within the numerical error) with an exponent $\mathcal{C}$,  characterising deviations from ODLRO in the thermodynamic limit, equal to $1/2$. For very large number of particles we therefore confirm the validity of Eq.~\eqref{universalresult}, independently of the external potential. It is also interesting to observe that not only 
Eq.~\eqref{universal_relation} is satisfied for the different power--law potentials analysed, but also that it is possible to work out predictions for $\beta$ and $\gamma$ as a function of $n$. 
For $\beta$ one can observe that, recalling the definition of length scale $\xi$ in Eq.~\eqref{def_xi}, the support of the dimensionless ground state natural orbit scales as
$\xi^{-1}$, \textit{i.e.}
\be
\label{xi_powerprediction}
\int d\eta \, \propto \, \frac{1}{\xi} \propto \hbar^{-\frac{1}{n+1}}\,.
\ee
Since $\int d\eta \propto N^{-2\beta}$ (see Sec.~\ref{sec3}) and in the semiclassical limit $\hbar \propto N^{-1}$, then we have
\be
N^{-2\beta}\,\propto\, N^{\frac{1}{n+1}}\,,
\ee
from which
\be
\label{guessing_betan}
\beta\,=\,-\frac{1}{2n +2}\,.
\ee
From the universal relation (\ref{universal_relation}) it also follows a prediction for $\gamma$:
\be
\label{guessing_gamman}
\gamma\,=\,\frac{n+3}{2(n+1)}\,.
\ee
The main results for the exponents $\mathcal{C}$, $\gamma$ and $\beta$ for different power--law potentials are reported in Tab.~\ref{table:finalresults}. Hence, the predictions (\ref{guessing_betan}) and (\ref{guessing_gamman}) are in excellent agreement with the numerical results.

Finally let's observe that the result $\gamma=1$ for $n_{\text{peak}}/\xi$ in the case of harmonic potential does not imply at all that in an experiment one would see a BEC, (\textit{i.e.} a macroscopic occupation of the lowest energy state). Indeed, also $\xi$ has a dependence on $N$. In experiments where $\tilde{\rho}(k)$ is measured, 
from Eq.~\eqref{xi_powerprediction} one would have
\be
n_{\text{peak}}\,\sim\, N^{1/2}\,.
\ee
The same behaviour is obtained for all values of $n$. This shows that for a TG gas the condensate {\it fraction} is $1/\sqrt{N}$ independently of the external trapping potential used to confine the system. 

\section{Conclusions}
\label{sec:Concl}
\noindent
In this paper we have studied the universal off--diagonal long--range order behaviour for a trapped Tonks--Girardeau gas at zero temperature. Firstly we have focused on the scaling of the largest eigenvalue $\lambda_0$ of the one--body density matrix of the gas with respect to its particle number $N$, defining the exponent $\mathcal{C}$ via the relation $\lambda_0 \sim \mathcal{B}N^{\mathcal{C}}$. For the one--dimensional homogeneous Tonks--Girardeau gas a well known result is $\mathcal{C}=1/2$. Here we have investigated the inhomogeneous case and we  have showed that $\mathcal{C}= 1/2$ actually characterises the hard--core system independently of the external trapping potential.
We also derived analytical predictions for the pre-factor $\mathcal{B}$. The field theoretical approach on which we relied shows clearly that the  large-$N$ asymptotic of the largest eigenvalue of the OBDM is the same varying the density and fixing the external potential or varying the external potential and fixing the density.

We have then defined the exponents $\gamma$ and $\beta$ of the scaling against $N$ of the dimensionless momentum distribution peak and the eigenfunction of the one--body density matrix relatives to $\lambda_0$, respectively. We have also defined a scaling length $\xi$, in terms of which we have introduced a dimensionless variable $\eta$ as $\eta\,=\,x/\xi$, further showing that $\xi$ scales with $N$ as $\xi\,\propto\,N^{2\beta}$ (the factor $2$ is introduced for convenience). The dimensionless ground--state natural orbital is then defined as 
$\hat{\varphi}_0(\eta)\,\equiv\,\varphi_0(x)\,\sqrt{\xi}\,\propto\,\xi^{1/2}$, due to the normalization condition of $\varphi_0(x)$. Therefore, as one inserts more particles into the system, the dimensionless natural orbital corresponding to $\lambda_0$ are wider, as expected. It then follows that $\hat{\varphi}_0(\eta)\,\sim\,N^{\beta}$. Another power--law scaling can be defined for the dimensionless momentum distribution peak $n_{\text{peak}}/\xi \sim N^{\gamma}$. Then we have showed that $\gamma + 2\beta = \mathcal{C}$. 

Confining the system in a power--law potential, $V(x) \propto x^{2 n}$, we were able to get analytical predictions for $\beta$ and $\gamma$. Using a semiclassical approximation approach we have found that $\beta\,=\,-1/(2n + 2)$ and $\gamma =(n+3)/[2(n+1)]$. 
We provided numerical checks for these predictions, using both a WKB approximation and exact numerical results. We have finally showed that it holds the following power--law scaling 
for the (dimensionful) momentum distribution peak: $n_{\text{peak}},\propto\,N^{1/2}$, valid for any external power--law potential. This is another universal property for a hard--core bosons  analogous to the one for the largest eigenvalue $\lambda_0$. The result for $n_{\text{peak}}$ is of interest for experiments since one has access to momentum distribution profiles, and therefore for a TG gas in a trap a condensate fraction of the order of $1/\sqrt{N}$ would be seen.

As a future work, it would be interesting to study the universal properties of the off--diagonal long--range order for a trapped Lieb--Liniger gas with finite coupling constant.

\vspace{3mm}
{\it Acknowledgements.} Discussions and useful correspondence with J. Dubail are gratefully acknowledged. 
GM and AT are grateful to the Erwin Schrödinger International Institute for 
Mathematics and Physics (ESI) in Wien for the kind hospitality during the 
programme ``Quantum Paths''. JV thanks SISSA and INFN for the kind hospitality during the final stage of this work.

\appendix
\section{Gauss Quadrature Method}
\label{app:A}
\noindent
The Gauss quadrature rule is a method with which one can estimate in terms of a finite sum an integral of a function $f(x)$ of the form
\be
\label{app_integral}
\int_a^b f(x) \,w(x)\,dx\,,
\ee
where $w(x)$ is some weight function. In the Gauss quadrature method the weights and nodes (points) where evaluating $f(x)$ are chosen in advance. This choice is based on the support of the integral in Eq.~\eqref{app_integral}. For example with the Gauss--Hermite quadrature one is able to compute integrals with $a\equiv -\infty$ , $b\equiv \infty$ and weight function $w(x)=e^{-x^2}$, in the following way
\be
\label{app_GHQ}
\int
f(x)\,e^{-x^2} \,dx \,\approx\,\sum_{i=1}^Z w_i\,f(\xi_i)\,,
\ee
where the $\xi_i$'s are the roots of the Hermite polynomial $H_Z(x)$, and (\ref{app_GHQ}) is exact for all polynomials $f(x)$ of degree less or equal than $2Z-1$. 

For the case of our interest $f(x)\equiv \rho(x,y)$. In order to recast (\ref{diagonalization_problem}) in the form of (\ref{app_GHQ}), we have to multiply and divide by $e^{-y^2}$. By choosing $x = \xi_k$, with $k=1,\dots,Z$, we have $Z$ equations of the form
\be
\label{app_integraleq}
\int
\left[\rho(\xi_k,y)\,\varphi_j(y)\,e^{y^2} \right] e^{-y^2}\,dy\,=\,\lambda_j\,\varphi_j(\xi_k)\,.
\ee
Using (\ref{app_GHQ}) we then have 
\be
\sum_{i=1}^Z w_i \,\rho(\xi_k,\xi_i)\,\varphi_j(\xi_i)\,e^{\xi_i^2}\,=\,\lambda_j\,\varphi_j(\xi_k)\,,
\ee
providing an eigenvalue equation for a $Z \times Z$ matrix $S$ with entries
\be
\label{app_Smatrix}
S_{k,i}\,=\,\rho(\xi_k,\xi_i)\,w_i\,e^{\xi_i^2}\,.
\ee
Accordingly, one has to diagonalize this finite dimensional matrix to obtain the occupation numbers $\lambda_j$ and the natural orbitals $\varphi_j(x)$. Of course, the larger is $Z$ and the better are the approximation results for the integrals. One has anyway to check whether increasing $Z$ the resulting value for the integral is converging. In the cases considered in the paper this condition was fulfilled and we used $Z$ ranging from $\approx 80$ to $\approx 170$. Moreover, one has a condition to check, that is
\be
\label{app_normcond}
\sum_{j=1}^Z \lambda_j \,=\,N\,,
\ee
with $N$ the number of particles in the system. If Eq.~\eqref{app_normcond} is not satisfied, then we have to increase $Z$.

If the support of the OBDM is compact, as in the case of the CFT limit of Sec.~\eqref{sub:pred_B}, one can rely on other quadrature scheme. For instance, the Gauss--Legendre quadrature method  can be applied to integrals having integration domain $[-1, 1]$ and gives
\be
\int_{-1}^{1} f(x)\,dx\,\approx\,\sum_{i=1}^Z w_i\,f(\xi_i)\,.
\ee
For the case of the half harmonic oscillator, one can use the same procedure but with different weights and nodes. 
The integration interval is $\left[0,\infty\right)$ and one has to apply the Gauss--Laguerre quadrature method
\be
\int_0^{\infty} f(x)\,e^{-x}\,dx\,\approx\,\sum_{i=1}^Z w_i\,f(\xi_i)\,.
\ee
The final form of the $Z \times Z$ matrix $S$ to diagonalize is then
\be
S_{k,i}\,=\,\rho(\xi_k,\xi_i)\,w_i\,e^{\xi_i}\,,
\ee
for $k,i = 1,\dots,Z$.

\end{document}